# A Georgi-Machacek Interpretation of the Associate Production of a Neutral Scalar with Mass around 151 GeV

François Richard[1]

Université Paris-Saclay, CNRS/IN2P3, IJCLab[2], 91405 Orsay, France
______________________________________________________________________

**Ongoing work presented at the ILC Workshop on Potential Experiments (ILCX2021), october 2021**

**Abstract**: *Following an investigation of several indications for new physics from LHC data, a detailed exploration of the Georgi-Machacek (GM) model was elaborated in a previous publication. This framework is used here to interpret the findings of Crivellin et al., which were obtained by combining the 2 photon mass spectrum from ATLAS and CMS, adding extra conditions, the most efficient one being the presence of >90 GeV missing transverse energy. The claim is a ~5 s.d. excess at 151 GeV. The GM model naturally explains this as coming from the cascade A(400)->H(151)Z, where H(151) is an **isosinglet** predicted by GM and where the accompanying signal comes from the decay of a Z boson into neutrino pairs. Satisfying the constraints from LHC measurements of the ZWW channel, which can receive a large contribution from this cascade, suggests that H(151) could dominantly decay into two light CP-odd scalars, following an ATLAS indication for h(125) decaying into bbμμ. A spectacular occurrence of signals is predicted at an e+e- collider operating up to 1 TeV, an energy sufficient to cover the whole scenario.*

## Introduction

In my interpretation of the various LHC indications [1] within the Georgi Machacek model, GM, there was an essential missing element: an isosinglet H, partner of h(125), as predicted within this model.

Since then, a new scalar has been convincingly observed at a mass of 151 GeV in its two photons decay mode [2]. This signal was not apparent in the genuine spectra but was obtained by requiring additional features like transverse missing energy, presence of two additional jets, leptons, b jets. This analysis combines ATLAS and CMS data. This is a "premiere" since this work, performed outside the two official collaborations, includes the integrated luminosities recorded by both experiment at 13 TeV.

---

[1] richard@lal.in2p3.fr
[2] Laboratoire de Physique des 2 Infinis Irène Joliot-Curie

The search of [2] originates from a detailed analysis of LHC data [3], which revealed various **topological excesses** comprising same sign leptons, leptons + b jets etc… These were interpreted as coming from cascades in which a heavy scalar decays into a lighter one with a mass of order 150 GeV. This prediction constitutes a strong point for this result, which does not suffer from the "look elsewhere" syndrome and can therefore claim a global evidence, at the **4.8 s.d. level, the highest so far observed at LHC, excluding the SM h(125).** Figure 1 summarizes this result where one also observes an additional ~2 s.d. contribution from the Zγ channel.

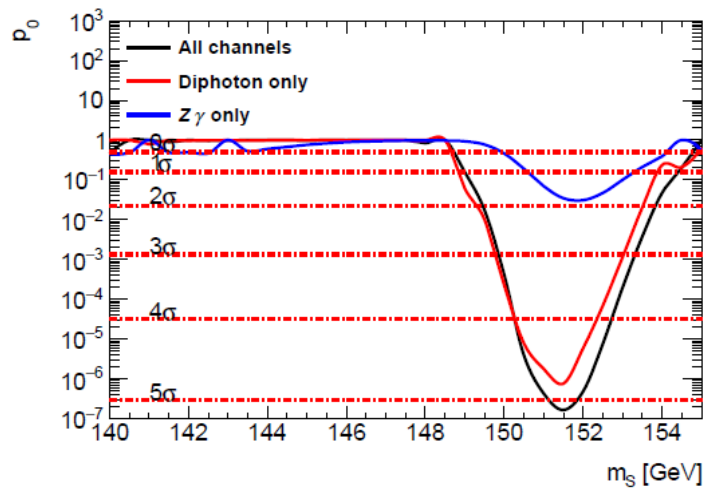

*Figure 1 : Combined local p-value as a function of the scalar mass from [2]*

The present note attempts to answer the following questions:

- Is this scalar the isosinglet H predicted by GM?
- Is it produced through the cascade predicted by this model **A(400)->HZ** ?
- Is GM able to describe quantitatively the LHC data ?

Recall that in [1] the main conclusion was that there are two heavy scalar candidates. One, a CP-even neutral scalar of a 5-plet, **H5(660),** identified in the ZZ mode selecting four leptons and, as expected, indicated in the VBF mode. Another one, a CP-odd neutral of a triplet, **A(400)**, identified in top pairs, tau pairs and Zh(125), the two latter being accompanied by b-jets. The Zh mode, which justifies the CP-odd interpretation, is most clearly seen by requesting an additional b-jet. A natural expectation is the process A->ZH(151) which would constitute a perfect candidate to explain the findings of [2]. The following picture summarizes this interpretation:

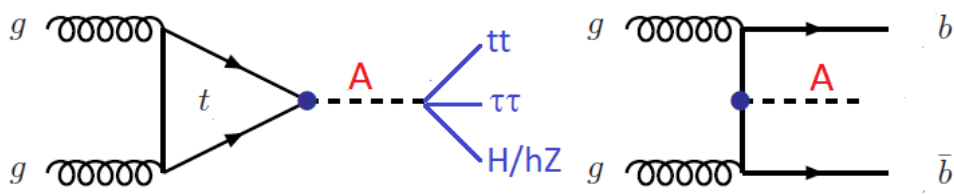

The first diagrams indicates that H and h are accompanied by a Z boson, which can provide a b-tag through Z->bb or a missing transverse energy through Z->$\nu\nu$. It has the highest cross section, several pb, within the GM model. The Yukawa coupling can be enhanced in an extended GM model, EGM, by a free factor $\zeta_t$ in the A2HDS scheme [4] by adding an extra doublet as explained in [1]. GM predicts that A(400)->H(151)+Z should be enhanced by a factor ~20 with respect to h(125)Z as shown in the following table from [1].

The second diagram is significant if the Yukawa coupling is multiplied by $\zeta_b$>10, within the **A2HDS** scheme [4]. This process is a good discovery process since it provides b-tagging for A->$\tau\tau$ and for A->hZ, as observed by ATLAS. Details about this EGM model can be found in the Appendix.

For the sake of simplicity, I will ignore the role of the second diagram for the H(151) interpretation since, as we shall see, the first diagram in the framework of the genuine GM model is able to explain the features observed by [2]. Not forgetting however that the second diagram with enhanced Abb coupling is needed to explain ATLAS observations of the hZ and $\tau\tau$ accompanied by b-jets and will also contribute to HZ production.

I also ignore the process gg->H(151), which has a cross section of ~1000 fb but offers no tagging opportunities.

In the GM model the Att coupling goes like $\tan\theta_H$~0.6, as shown in the table below.

| Type | coupling/SM | $s\alpha$=-0.15 sH=0.5 |
|---|---|---|
| h(125)WW/ZZ | $c\alpha cH - 1.63 s\alpha sH$ | 0.99 |
| H(151)WW/ZZ | $s\alpha cH + 1.63 c\alpha sH$ | 0.68 |
| Att,bb,$\tau\tau$ | tanH | 0.58 |
| h(125)tt,bb | $c\alpha$/cH | 1.14 |
| H(151)tt,bb | $s\alpha$/cH | 0.17 |
| Zh(125)A | $1.63(s\alpha cH+0.6c\alpha sH)$ | 0.28 |
| ZH(151)A | $1.63(c\alpha cH-0.6s\alpha sH)$ | 1.48 |
| Wh(125)H3$^+$ | $1.63(s\alpha cH+0.6c\alpha sH)$ | 0.28 |
| WH(151)H3$^+$ | $1.63(c\alpha cH-0.6s\alpha sH)$ | 1.48 |

This table predicts that H(151) has a small coupling to fermions and a reduced coupling to W/Z, meaning that H should have a narrow total width, ~5 MeV. As for the SM h(125), this feature is essential since it opens the possibility of revealing new physics, visible or invisible. On the later, the search conducted for h(125) can be easily be translated to H(151) and allows to set an upper limit of:

**BR(H(125)->invisible)<40%**

The two photon decay search is a priori favoured by the absence of cancellation between W and top contributions. In GM there is a rich spectrum of charged scalars, which will also contribute to this BR in a way that is discussed in the Appendix. One can just provide a preliminary evaluation:

$$\Gamma(H(151) \to \gamma\gamma) \sim 0.025 \text{ MeV}$$

The coupling to tt being also suppressed, the production cross section for gg->H is 50 times smaller than for h(125), hence not observable in the standard ZZ and WW analyses performed at LHC. The main source of H(151) comes from the cascades A->HZ originating from the two diagrams previously discussed. This feature is essential since it provides the possibility of self-tagging of the H decays by the accompanying Z, hence the success of [2]. Note, in passing, that contrary to [2], the GM interpretation does not require introducing a new particle decaying invisibly.

GM predicts a large cross section for the process gg->A->HZ:

$$\sigma(gg \to A \to HZ) = 6000 \, BR(A \to HZ) = 3300 \text{fb}$$

The GM model predicts that H decays in 90% of the cases into WW. This gives a cross section for ZWW in flagrant contradiction with the measurements of CMS, which [5] tell us that this cross section should be below 400 fb.

Should we therefore drop the GM interpretation of [2]? By no means! Again, the very low width predicts that any new BSM channel can easily dominate over the SM decays. This naturally introduces the well-known scenario already explored by ATLAS [6] where there could be a decay **H->aa**, where a is a light CP-odd scalar.

In [6], it was found that there could be such a mechanism for h(125) with $m_a=52$ GeV and $BR(h \to aa \to bb\mu\mu) \sim 2 \, 10^{-4}$. The $bb\mu\mu$ channel has been used since it allows the easiest identification in the LHC environment but one expects that the dominant channel should be bbbb and, assuming that the Yukawa coupling of a are proportional to the fermion masses, one expects a BR(H->aa)~25%, which would correspond to a width of about 1 MeV.

Again one might think that this result applied to H(151) does not allow to solve the **ZWW paradox,** which requires an additional order of magnitude of the total width. To understand this, one may simply invoke the fact that in GM the non doublet content of h(125) is suppressed by a mixing angle which has to be small to insure that h behaves dominantly as a doublet. If the coupling haa proceeds through the triplet component, one can understand that H(151) can have enhanced coupling to aa. I will show in section I that this idea, still very speculative, works beautifully to solve the ZWW paradox.

This low mass for H allows fulfilling **the unitary constraints** of GM as will be argued in section II, while this would not be the case for a heavier scalar. One therefore concludes that H(151) fulfils the criteria to be identified as the missing neutral isosinglet of the GM model.

For what concerns future e+e- colliders, the process **e+e- ->ZH(151)** is produced with half the cross section of a genuine SM Higgs and is fully accessible at ECM=250 GeV, a centre of mass energy provided by all projects under study, which could therefore act as factories for the two scalars h(125) and H(151).

# Section I Quantitative interpretation of the 151 GeV signal

In [1], the following diagram was summarizing the particle content of the GM model and the transitions expected between 5-plets and 3-plets and 3-plets and singlets. In this work, I will identify the scalar found in [2] with the missing singlet H. In green are shown identified particles and processes.

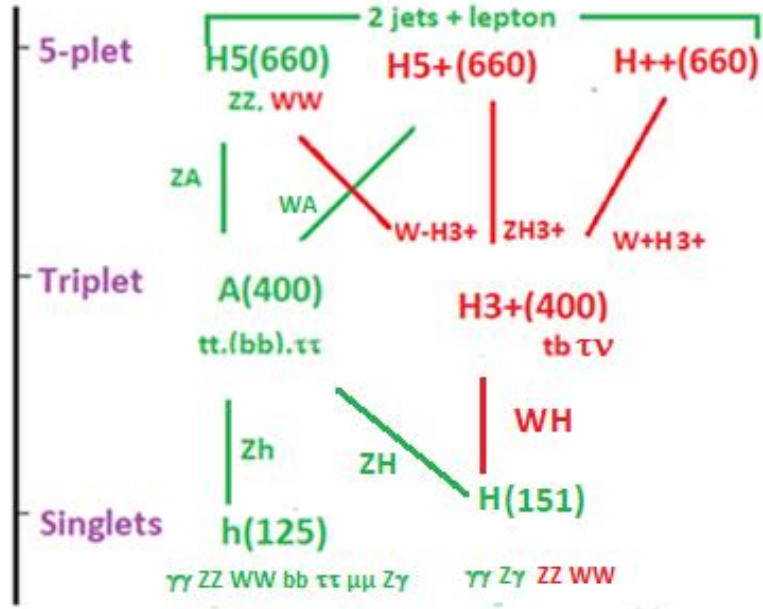

An obvious comment is the absence of charged Higgs indications.

## Section I.1 Decay widths of A(400) and H(151)

For simplicity, I will assume GM, which gives:

| $\Gamma$Atot GeV | $\Gamma$tt GeV | $\Gamma$hZ GeV | $\Gamma$HZ GeV |
|---|---|---|---|
| 14.8 | 6.4 | 0.4 | 8 |

With the exception of the decay into $2\gamma$ and $Z\gamma$, which require a knowledge of loop effects (see Appendix), the H(151) widths are calculable, once the masses are fixed, in term of two parameters:

- The mixing angle $\alpha$ affecting h and H, which needs to be small to preserve h SM properties
- A mixing angle $\theta_H$ which fixes the vacuum expectation of the triplets

These two parameters were chosen as $\sin\alpha=-0.15$ and $\sin\theta_H =0.5$ in [1]. This choice seems critical for what concerns $\Gamma$hZ, for which occurs an accidental cancellation. To see this, one can instead assume $\sin\alpha=-0.1$ and $\sin\theta_H =0.65$, which gives $\Gamma$hZ = 1.3 GeV, while $\Gamma$HZ remains almost unchanged. This aspect only matters for interpreting the evidence for A->hZ observed by ATLAS, which is irrelevant in the present discussion. Therefore, I will keep these two GM parameters unchanged.

With these parameters, GM gives $\sigma$(gg->A->HZ)=6000BR(A->HZ)=3300fb.

The boson H decays primarily into WW and ZZ as shown in the table below. The low mixing angle $\alpha$ suppresses the fermion couplings .

| $\Gamma$Htot MeV | $\Gamma$WW MeV | $\Gamma$ZZ MeV |
|---|---|---|
| 5 | 4.5 | 0.5 |

$\Gamma$ **(H->2γ)=0.025 MeV,** if one takes into account the W loop (the top exchange can be neglected for an isosinget) and the charged scalar loops (see the Appendix).

From the LHC result for BRh->invisible), which depends on the coupling of h/H to ZZ, one can deduce that BR(H->inv)<40%. This clearly says that invisible decays cannot solve the ZWW paradox mentioned in the introduction.

Again this description operates assuming that there are no BSM decay modes of H(151).

## Section I.2 The ZWW paradox

From the previous section, one finds that the process gg->AZ->ZWW has a ~3000 fb cross section. This result appears in violent disagreement with the following CMS results summarized below [5]:

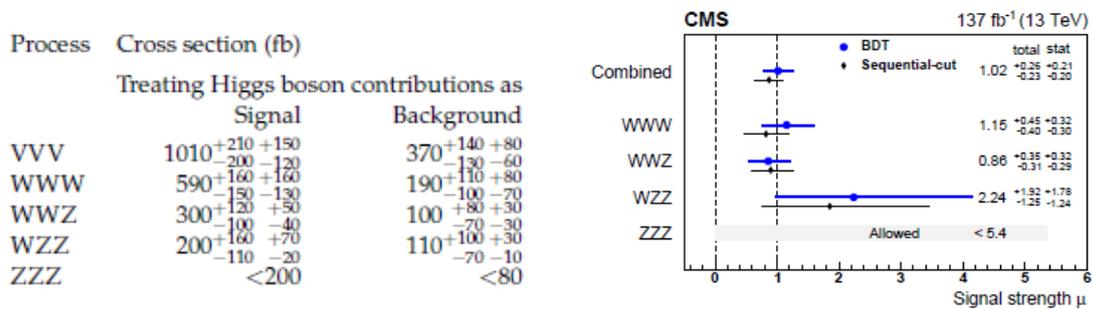

From this one can set a 2 s.d. bound on BSM contributions to ZWW:

$$\Delta\sigma(ZWW)<400 \text{ fb}$$

This result clearly requires that the BR of H into WW be reduced by an order of magnitude. This seems feasible, as discussed in the next section.

## Section I.3 The case for h/H->2a

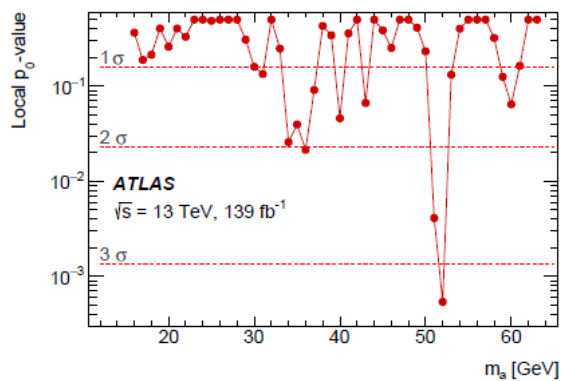

*Figure 2 : The local $p_0$–values versus the mass hypothesis for the search of h->aa in ATLAS.*

Various theoretical arguments recalled in [5] motivate the existence of light CP-odd scalars and the occurrence of the process h->aa. This line of reasoning offers way to solve the ZWW paradox by assuming that H mostly decays into aa, where a is a CP-odd scalar.

ATLAS [6] has carried a search for h->aa->bbμμ and observed an indication for a mass peak in the μμ system with ma~52 GeV, as illustrated by figure 2.

Admittedly, this effect has a low significance, 3.3 s.d. local and 1.7 s.d. global. Assuming that it is real, I will show that it offers a solution to the ZWW paradox.

The final state bbµµ is selected as it gives the best visibility at LHC, but it represent a minute fraction of the expected final states dominated by bbbb, bbττ, ττττ, bbcc etc... One expects that BR(bbµµ)/BR(bbbb)~$2m^2µ/3m^2b$=1/1300, where I took into account the colour factor 3 and the combinatorial factor 2 in favour of bbµµ. The estimated branching ratio BR(h->bbµµ)~210-4 multiplied by 1300, gives a BR(h->2a)~26% or Γ(h->2a)~1 MeV.

As recalled in the introduction, H and h may behave quite differently, given that h comes dominantly from the doublet of GM and H from the triplet. Within GM, the mixing suppression is sin$\alpha$~-0.15. If the coupling to the CP-odd scalars goes through the triplet component, one expects:

$$\Gamma(H->aa) \sim \Gamma(h->2a)/\sin^2\alpha \sim 44 \text{ MeV}$$

This would mean that the BR(H->WW) would be divided by 10, predicting σ (A->ZWW)~300 fb, which can be accommodated by the CMS results.

Another concern is whether the presence of this new source of bbµµ can be observed in the ATLAS analysis? Recall that the cross section for gg->AH is about 3000 fb hence a cross section of ~2.5 fb for the bbµµ mode, which is to be compared to the 10 fb cross section observed by ATLAS for h->bbµµ, therefore almost negligible.

In conclusion, one can build up a scenario, which provides an explanation of the ZWW paradox. This precise scenario remains purely **speculative**. What really matters is to assume that H(151) is decaying in one or several **BSM modes** in 90% of the cases and we know that these modes are **not all invisible**.

## Section I.4 The final interpretation

The following table summarizes the results achieved so far:

| ΓHtot MeV | ΓWW MeV | ΓZZ MeV | Γγγ MeV | Γaa MeV | Γinv MeV |
|---|---|---|---|---|---|
| 50 | 4.5 | 0.5 | 0.025 | 45 | <20 |

From this, one can estimate the cross sections for the two processes:

**gg->HZ->2γ+ETmiss and 4ℓ**

For the first process, one predicts:

BR(Z->νν)σ(A->HZ)BR(H->2γ)=0.2*3000(0.025/50)=0.3 fb, instead of 0.42+-0.13 from [2]

This cross section comes from counting the number γγ candidates after subtracting the background, only taking into account the integrated luminosity and the two photon reconstruction efficiency ~60%, without an attempt to include the effect of the ETmiss selection, which is model dependent.

For the second process, one predicts:

σ(4ℓ)=σ(A->HZ)BR(H->ZZ)BR(ZZ->4ℓ)= 3000(0.5/50)0.07²=0.15fb instead of <0.28 fb from [2]

Concerning the invisible width, the upper limit reached at LHC for h(125) can be interpreted in terms of H(151) by saying BR(inv)<40%. This means <20 MeV, to be compared to the limit <1 MeV for h(125). This allows for a much larger contribution of invisible decays for H(151).

# Section II Unitary constraints

In this section, I will indicate how the GM model allows constraining the mass of H(151). As already mentioned in [1], unitarity arguments put stringent upper limits on the masses of the 5-plet, triplet and singlet of order 700 GeV. Knowing m5 and m3 from the data, is it possible to predict the mass of the singlet H ?

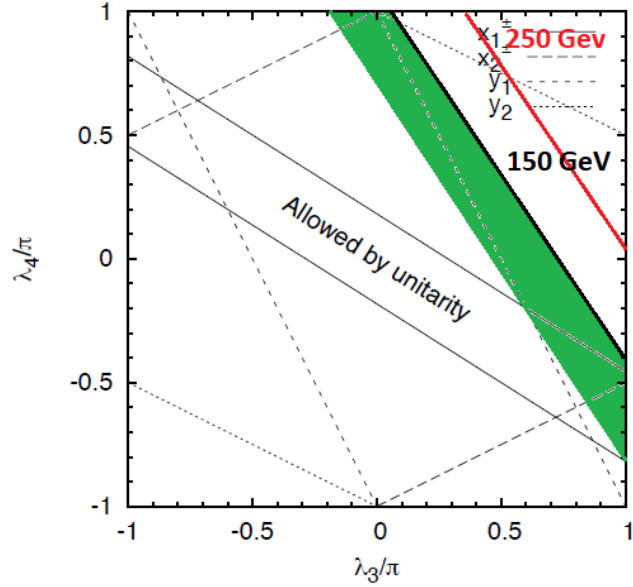

Figure 3 : Regions allowed by unitarity for the couplings $\lambda 3/\pi$ and $\lambda 4/\pi$ of the GM potential. The thick black line corresponds to the 150 GeV solution for an isosinglet, the red one to mH=250 GeV. The green area gives the lower part of the uncertainty on the GM parameters.

In the Appendix are given the detailed formulae for this derivation, in the convention of [7]. The main equation needed for this derivation reads:

$$8(3\lambda 3+2\lambda 4)v^2\chi = m_5^2 + 2(m_H^2\cos\alpha^2 + m_h^2\sin\alpha^2) - 3\cos\theta_H^2 m_3^2$$

where v=246 GeV and where m3 and m5 are deduced from LHC indications, m5=660±20 GeV and m3=420±20 GeV.

Note that this equation is independent of the two other parameters, M1 and M2, which enter in the GM scalar potential (see Appendix).

From this equation, one can draw in the plane $\lambda 3/\pi$ and $\lambda 4/\pi$ straight lines for each value of mH. Figure 3 shows that the line mH=150 GeV is compatible, within errors, to the unitarity band, while, for instance, the line 250 GeV is excluded by this criterion. This result shows that the allowed domain for mH is relatively narrow, which comforts our interpretation of H as a valid **isosinglet** candidate.

Note also that the preferred values for $\lambda 3/\pi$ is positive and between 1/2 and 4/5, the later limit comes from unitarity. $\lambda 4/\pi$ should be negative and above the unitarity limit -16/25.

Note also that this method does not allow to set a limit against mH<150 GeV, that is against the CMS 96 GeV candidate.

A similar strategy can be followed for $\lambda 2$ and $\lambda 5$, which, after eliminating the parameter M1, reads:

$$4\lambda 2 - \lambda 5 = \sin 2\alpha (m_H^2 - m_h^2)/2\sqrt{3}v_\chi v_\phi + 2m_3^2/v^2$$

Together with the combination of LHC results on h->2γ, which gives ϰγγ=**1.11+0.10-0.09** (PDG 2020), one sees that it is already possible to have a precise determination of λ2 and λ5.

A full derivation of the parameters of the GM potential is given in the Appendix, which allows computing BR(H->2γ) .

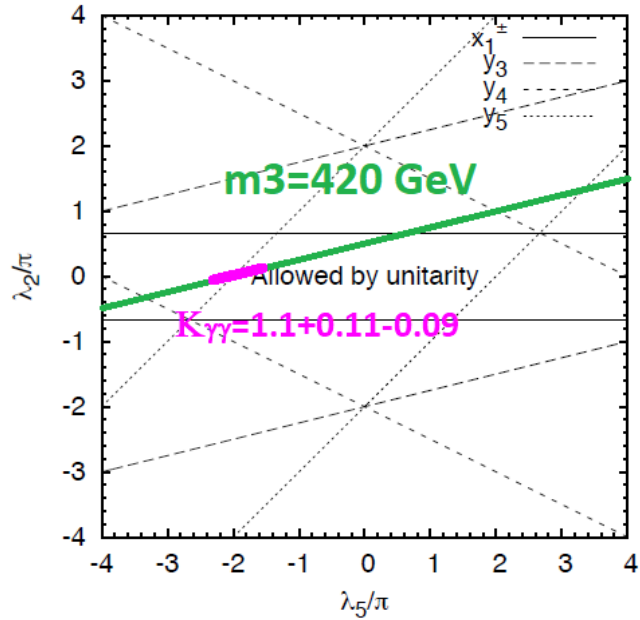

*Figure 4 : Area allowed by unitarity for the couplings λ2/π and λ5/π of the GM potential. The green line corresponds to an estimate deduced from the LHC indications. The magenta line gives the region allowed by the LHC measurements of h(125)->2γ.*

In conclusion, this analysis also comforts the interpretation of H(151) as a GM isosinglet.

# Section III Expectations at lepton colliders

## III.1 Annihilation process

Figure 5 shows the cross sections expected for the SM Higgs boson and the three lowest mass scalars indicated by LHC data. It is remarkable that the H(151)Z cross section is similar to the SM cross section while it is suppressed by an order of magnitude at LHC with respect to h(125).

In the three cases, given the high rates, it will be possible to achieve "exquisite" accuracies on the parameters of these scalars, as for h(125): for the **total width**, the **invisible width**, the **couplings to vector bosons and fermions.**

For H(151), one has an indication that 90% of its decay modes go through a pair of CP-odd scalars, **H->aa**, with ma=52 GeV. So far, LHC has only observed the bbμμ mode, ~0.1% of these decays. The channel HZ offers the possibility to have a clean selection of the **a(52)** decays and to measure their branching ratios. Setting a tight upper limit or, eventually, measuring the invisible width of H will also be of great interest.

Separating hZ from HZ final states can be easily achieved by using the recoil mass technique with Z->ℓ+ℓ- final states. Using the hadronic decays would increase the efficiency by an order of magnitude. This possibility deserves to be investigated in detail and will constitute an important benchmark for the various detectors under consideration.

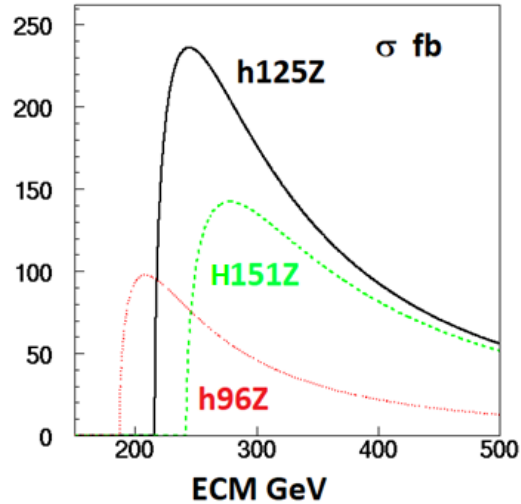

*Figure 5 : Predicted e+e- cross sections in fb vs. the centre of mass energy ECM for the SM Higgs, the isosinglet H(151) and the scalar h(96) indicated by CMS and LEP2 data.*

Interference effects between h(125) and H(150) should exist but be easy to control since the dominant final states are different: bb for h(125), aa for H(151).

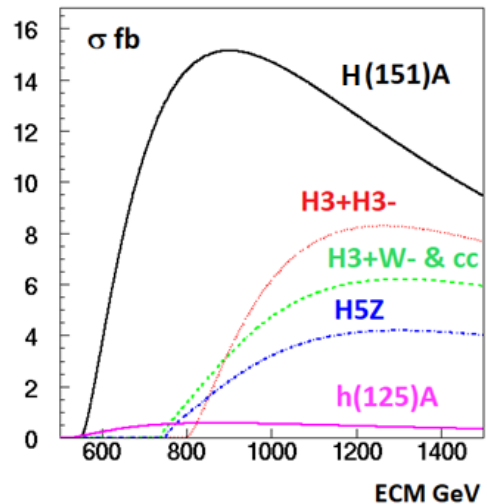

*Figure 6 : Predicted e+e- cross sections vs. ECM for the various processes expected within the GM model.*

Figure 6 shows the cross sections expected for the heavier scalars. These cross sections are much lower, implying a challenge for future e+e- colliders. On top of that, final states are much more complex, a challenge for the detectors, requiring a **full solid angle coverage**, in particular for **b-jet tagging**, a weakness of most solenoidal systems built so far. A full study, with realistic simulations of the various channels and a careful evaluation of efficiencies and backgrounds is therefore highly desirable to evaluate the performances of the various detectors on the market.

## III.2 Case for an e-e- collider

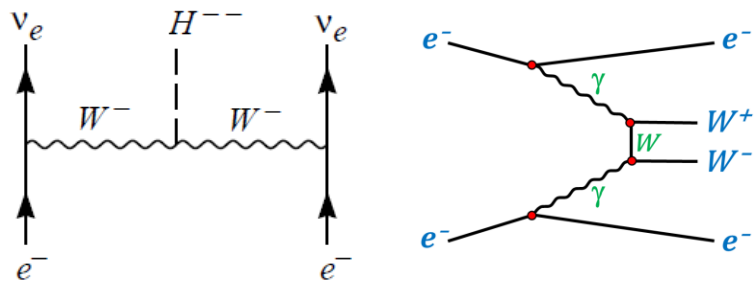

The process e+e→ H⁺⁺H⁻⁻ requires reaching ~1.5 TeV. An alternate is to consider VBF single H⁻⁻ production at 1 TeV, the process shown above:

- One of the largest backgrounds comes from the second diagram and is easily removed by appropriate cuts, e.g. by requiring a large missing transverse momentum of the final state and no spectator e+/e- [8]

- The decay mode **H⁻⁻(660)->H3⁻(400)W⁻** with **H3⁻ ->tb**, is easily separated from most backgrounds using **b-tagging** and applying a multijet selection

Furthermore:

- $\mathcal{L}$e-e-~70%$\mathcal{L}$e+e- seems feasible [9]
- The cross section for this process reaches 3.5 fb at 1 TeV for $\sin\theta H=0.5$
- With polarized beams, 80% for e-, one can achieve a luminosity gain, 1.8², for the W-W-luminosity, hence a cross section above 10 fb
- Switching from e+e- to e-e- will require changing polarity of many magnets, which seems manageable if anticipated
- An ERL scheme [10] seems feasible

## IV Precision measurements

SM couplings of h(125) are modified within the GM model as shown by the table of couplings given in the introduction. At LHC, where the total width cannot be measured, precision accuracies need to be taken with a grain of salt. Nevertheless, it is fair to say that the retained solution significantly differs from the SM predictions and offer good prospects for HL-LHC. A prominent example is the hbb coupling.

In the Appendix the loop contributions from GM charged Higgs to h(125)->2γ are given and it is shown that one can easily generate effects at the 10% level since the SM process itself is at the loop level. Less trivial will be the case of hZZ, but recall that the measurement from e+e- factories is at the few per mill level.

A completely new domain opens up if the appearance of h->aa and H->aa are confirmed. For h, a recoil mass technique will allow to isolate the prominent decay mode into bbbb(~25%). H->bbbb will be the dominant decay mode and can hope to observe bbττ, which could give up to 20% of the decay modes.

## Section V Prospects at LHC

The GM interpretation obviously offers a promising harvest of discoveries for HL-LHC. Is there more to it? An answer to this question is most probably YES. Recall that in [1] it was shown that the fermion couplings to A(400) were requesting an additional doublet, implying additional charged scalar H2+, neutral scalar H2 and pseudoscalar A2, where the index 2 refers to a doublet component. This brings more complexity in the definition of the GM potential and requires an input from theory to define rigorously this extended model. The physical states are presumably mixing between the doublet and the triplet component of GM. In particular, as was done in section II, one can hope to use **unitarity** to constrain the masses of these new scalars. Naively, one can think that since A3 and A2 should mix to produce the right pattern for fermions of A(400), their masses should lie nearby.

The experimental status and interpretation of **h(96)** still need to be clarified. Is it simply an additional isosinglet which, in particular, is expected to mix with h(125)? Or a Radion [11]? How can we distinguish between the two hypotheses?

One of the most striking conclusions reached in this paper is that both h(125) and H(151) could frequently decay into bbbb final states. For h(125) one expects a LHC cross section ~10 pb and for H(151) ~3 pb with the opportunity to use an additional tagging. A feasibility study of such an analysis is desirable.

One may also wonder why Nature has been so generous to add this large family of scalars to the abundant family of fermions. Since the fermionic sector is, by itself, unable to explain the primordial **electroweak transition and the baryogenesis process**, one is tempted to attribute to these scalars this mission. In particular measuring **CP violation** in this sector will be a capital but difficult mission ascribed to e+e- colliders. It is however fair to say that the GM scalar potential is not offering explicit CP-violation, meaning that GM will require further extensions.

More generally, GM like the SM are just effective theories lacking an UV completion. In this respect, it is intriguing that in [12] a phenomenology with triplets emerges naturally from a more ambitious approach towards an UV complete theory. Other particles like heavy top partners, are also predicted.

## Summary and conclusion

A **consistent interpretation** of the 151 GeV candidate found by [2] is provided within the GM model. More than the actual statistical significance of this signal, always disputable, this mere fact contributes to increase our confidence in this finding. A simple cascade mechanism, A->HZ, is proposed to explain [2], without arbitrarily introducing extra invisible particles.

This evidence for a singlet complements the CP-odd triplet A(400) and of a CP even H5(660) 5-plet. Given its low mass, H(151) opens a wide range of transitions, which will be exploited at HL-LHC or, even before, with Run3 and/or by combining the Run2 data of the two collaborations as was done in [2].

A natural consequence of LHC observations is that H(151) decays dominantly into BSM modes, mostly visible, possibly **H->aa** where a(52) is a light pseudo-scalar suggested by an indication from ATLAS.

The present interpretation reconciles the **topological approach** [3] and the **spectroscopic approaches** [1] and [2], providing a reliable consolidation of the various **evidences for BSM physics at LHC** discussed in these three references.

From LHC experiments one expects a full confirmation *of H(151), A(400), H5(660)* and, possibly, S(96). An exploration of **WWW and ZWW** final states is desirable since these channels can receive large contribution from the GM cascades. Discovering some of the numerous **charged scalars** predicted by this theory is an essential task for HL-LHC.

Needless to repeat that an e+e- machine **reaching one TeV** will provide an ideal tool to investigate this new zoology and achieve very precise measurements of the parameters of these particles and their CPV properties. Of particular interest is the possibility offered by the HZ channel to confirm the existence of H->aa and the decay properties of a(52). This narrow resonance also offers the opportunity to measure precisely its invisible decay modes. The doubly charged boson $H5^{--}$ predicted by GM can be singly produced with a 1 TeV collider operating in the e-e- mode with a luminosity similar to that of e+e-.

There are **demanding** extra **requirements on detectors**, calling for an almost perfect solid angle coverage, indispensable for efficiently reconstructing events with **up to 10 jets**, in clear distinction to what is presently achievable for a Higgs factory tuned for the simple hZ topology. This coverage includes b-quark identification and appears very challenging. Reaching a **high reconstruction efficiency** for these complex modes is mandatory given the cross-sections ~10 fb.

An important input from this work is an access to the parameters of the **complex scalar GM potential.** The present solution appears to satisfy the **unitarity constraints** in a non-trivial way, which would not be the case if the H particle had been heavier, say with a mass above 250 GeV. Using the process h(125)->2$\gamma$ measured at LHC, it is shown that even with the present accuracy, one can already tightly constraint some parameters of the Higgs potential. It is shown that a precise measurement of **BR(H->2$\gamma$)**, provided by LHC, gives indirectly access to the loop contributions coming from the rich sector of the **charged scalars** of GM.

More generally, precision measurement will also play a critical role in providing the final assessment of the GM interpretation. As an example, this analysis predicts a ~15% deviations on the h(125)bb coupling, already observable at HL-LHC.

If such GM indications were confirmed, the **international initiatives** towards future e+e- colliders should be greatly boosted toward using the linear collider approach to reach an appropriate centre of mass energy.

**Acknowledgements**: *I am grateful to my colleagues from IJCLab M. Davier, A. Falkowski, A. Le Yaouanc, R. Poeschl, D. Zerwas and Z. Zhang for kindly encouraging this work. I also thank G. Moultaka from Université de Montpellier L2C, for providing his expertise on the Georgi-Machacek model. More recently, I had the pleasure to share several ideas with the authors of [16], and would like to show my appreciation of their very helpful responses, hoping to develop further our collaboration.*

# APPENDIX

## I. The general framework

Since the conventions used for the GM model greatly vary from author to author, I will recall below the ones used in the present analysis and derive some results deduced from the various LHC indications.

GM is constituted by one doublet $\phi$ and two triplets X [7]:

$$\Phi = \begin{pmatrix} \phi^{0*} & \phi^+ \\ -\phi^{+*} & \phi^0 \end{pmatrix},$$

$$X = \begin{pmatrix} \chi^{0*} & \xi^+ & \chi^{++} \\ -\chi^{+*} & \xi^0 & \chi^+ \\ \chi^{++*} & -\xi^{+*} & \chi^0 \end{pmatrix}$$

forming the following triplet H3 and 5-plet of physical states:

$$H_5^{++} = \chi^{++},$$
$$H_5^+ = \frac{(\chi^+ - \xi^+)}{\sqrt{2}},$$
$$H_5^0 = \sqrt{\frac{2}{3}}\xi^0 - \sqrt{\frac{1}{3}}\chi^{0,r},$$
$$H_3^+ = -s_H \phi^+ + c_H \frac{(\chi^+ + \xi^+)}{\sqrt{2}},$$
$$H_3^0 = -s_H \phi^{0,i} + c_H \chi^{0,i}.$$

Triplets are a mixture of X and ϕ with a mixing governed by θ_H, while α governs the singlet mixing.

There are also two singlets H1 and H1', with the following composition:

$$H_1^0 = \phi^{0,r},$$
$$H_1^{0\prime} = \sqrt{\frac{1}{3}}\xi^0 + \sqrt{\frac{2}{3}}\chi^{0,r}.$$

meaning that H1 comes from the doublet ϕ.

The physical states, h(125) and H(151) are combinations of these GM singlets with a mixing angle α:

$$h = \cos\alpha\, H_1^0 - \sin\alpha\, H_1^{0\prime},$$
$$H = \sin\alpha\, H_1^0 + \cos\alpha\, H_1^{0\prime}.$$

This mixing angle has to be small to avoid altering the doublet properties of the SM h(125). From [1]:

sinα=-0.15  vϕ=213 GeV  vχ=43.5 GeV.

vϕ and vχ, the two vacuum expectations for the doublet and the two triplets are related to the SM vacuum expectation: v²= vϕ²+8vχ². Redundantly one usually defines a mixing angle θ_H:

$$c_H \equiv \cos\theta_H = \frac{v_\phi}{v}, \qquad s_H \equiv \sin\theta_H = \frac{2\sqrt{2}\,v_\chi}{v}$$

with the following choice taken in [1]: sinθ_H=0.5.

The triplet and 5-plet masses are m3 and m5. From LHC indications, m3 and m5 are determined, with some uncertainties. The CP-odd meson A has a mass difficult to ascertain from the top pair analysis, given interferences with the QCD background. The channel Abb->Zhbb, which gives the best evidence, prefers 440 GeV, while other indications cluster at ~400 GeV. Tentatively, I assume that m3=420±20 GeV. H5(660) is a wide resonance and, accordingly, I will assume that m5=660±20 GeV.

The scalar potential reads:

$$V(\Phi, X) = \frac{\mu_2^2}{2}\mathrm{Tr}(\Phi^\dagger\Phi) + \frac{\mu_3^2}{2}\mathrm{Tr}(X^\dagger X) + \lambda_1[\mathrm{Tr}(\Phi^\dagger\Phi)]^2 + \lambda_2\mathrm{Tr}(\Phi^\dagger\Phi)\mathrm{Tr}(X^\dagger X)$$
$$+ \lambda_3\mathrm{Tr}(X^\dagger X X^\dagger X) + \lambda_4[\mathrm{Tr}(X^\dagger X)]^2 - \lambda_5\mathrm{Tr}(\Phi^\dagger\tau^a\Phi\tau^b)\mathrm{Tr}(X^\dagger t^a X t^b)$$
$$- M_1\mathrm{Tr}(\Phi^\dagger\tau^a\Phi\tau^b)(UXU^\dagger)_{ab} - M_2\mathrm{Tr}(X^\dagger t^a X t^b)(UXU^\dagger)_{ab}.$$

To extract the various parameters, I have used the following expressions:

$$m_5^2 = \frac{M_1}{4v_\chi}v_\phi^2 + 12M_2 v_\chi + \frac{3}{2}\lambda_5 v_\phi^2 + 8\lambda_3 v_\chi^2,$$
$$m_3^2 = \frac{M_1}{4v_\chi}(v_\phi^2 + 8v_\chi^2) + \frac{\lambda_5}{2}(v_\phi^2 + 8v_\chi^2) = \left(\frac{M_1}{4v_\chi} + \frac{\lambda_5}{2}\right)v^2.$$

$$\mathcal{M}_{11}^2 = 8\lambda_1 v_\phi^2,$$
$$\mathcal{M}_{12}^2 = \frac{\sqrt{3}}{2}v_\phi\left[-M_1 + 4(2\lambda_2 - \lambda_5)v_\chi\right],$$
$$\mathcal{M}_{22}^2 = \frac{M_1 v_\phi^2}{4v_\chi} - 6M_2 v_\chi + 8(\lambda_3 + 3\lambda_4)v_\chi^2.$$

The mixing angle is fixed by

$$\sin 2\alpha = \frac{2\mathcal{M}_{12}^2}{m_H^2 - m_h^2},$$
$$\cos 2\alpha = \frac{\mathcal{M}_{22}^2 - \mathcal{M}_{11}^2}{m_H^2 - m_h^2},$$

with the masses given by

$$m_{h,H}^2 = \frac{1}{2}\left[\mathcal{M}_{11}^2 + \mathcal{M}_{22}^2 \mp \sqrt{(\mathcal{M}_{11}^2 - \mathcal{M}_{22}^2)^2 + 4(\mathcal{M}_{12}^2)^2}\right].$$

## II.     Unitarity constraints

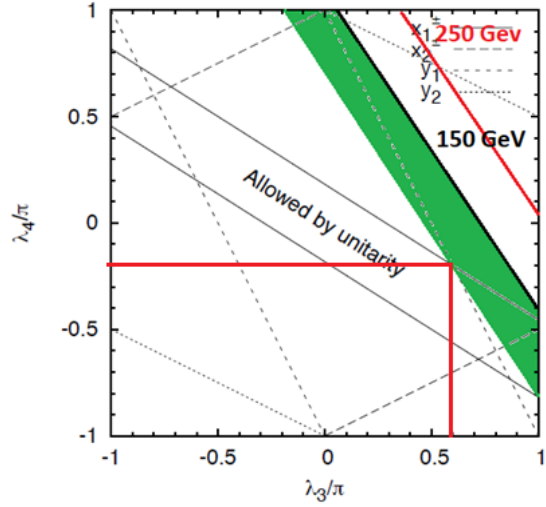

*Figure 7: Same as figure 3, showing the retained solution used to compute the GM parameters.*

Using above equations, one can eliminate the parameters M1 and M2, writing that:

$$4\lambda 2 - \lambda 5 = \sin 2\alpha (mH^2 - mh^2)/(2\sqrt{3}v\chi v\phi) + 2m^2 3/v^2$$

Similarly, one can write:

$$8(3\lambda 3 + 2\lambda 4)v^2\chi = m5^2 + 2(mH^2 \cos\alpha^2 + mh^2 \sin\alpha^2) - 3\cos\theta H^2 m3^2$$

Knowing the parameters of the second member of these equalities, one can draw straight lines in the planes λ2/λ5 and λ3/λ4 as was done in the main text. This allows deciding if these parameters fulfil the unitarity requirements defined in [7]. It turns out that they barely do it for λ3/λ4, which allows fixing these values in a narrow window.

## III.     Determination of the parameters of the scalar potential

To determine λ2/λ5, one can use the measured value of κγγ for the process h->γγ given by a recent PDG averaging of ATLAS and CMS:

$$\kappa\gamma\gamma = 1.11 + 0.11 - 0.09$$

This measurement depends on λ2/λ5, through the charged scalar loops of GM as shown from the formulae of reference [13]:

$$C_{H_3^+H_3^{-*}h} = \frac{1}{\sqrt{3}v^2}\left\{\sqrt{3}c_\alpha\left[(4\lambda_2-\lambda_5)v_\phi^3+8(8\lambda_1+\lambda_5)v_\phi v_\chi^2+4M_1v_\phi v_\chi\right]\right.$$
$$\left.-s_\alpha\left[8(\lambda_3+3\lambda_4+\lambda_5)v_\phi^2 v_\chi+16(6\lambda_2+\lambda_5)v_\chi^3+4M_1v_\chi^2-6M_2v_\phi^2\right]\right\}$$
$$C_{H_3^+H_3^{-*}H} = \frac{1}{\sqrt{3}v^2}\left\{\sqrt{3}s_\alpha\left[(4\lambda_2-\lambda_5)v_\phi^3+8(8\lambda_1+\lambda_5)v_\phi v_\chi^2+4M_1v_\phi v_\chi\right]\right.$$
$$\left.+c_\alpha\left[8(\lambda_3+3\lambda_4+\lambda_5)v_\phi^2 v_\chi+16(6\lambda_2+\lambda_5)v_\chi^3+4M_1v_\chi^2-6M_2v_\phi^2\right]\right\}$$
$$C_{H_5^+H_5^{-*}h} = C_{H_5^{++}H_5^{--*}h} = c_\alpha\left[(4\lambda_2+\lambda_5)v_\phi\right]-\sqrt{3}s_\alpha\left[8(\lambda_3+\lambda_4)v_\chi+2M_2\right],$$
$$C_{H_5^+H_5^{-*}H} = C_{H_5^{++}H_5^{--*}H} = s_\alpha\left[(4\lambda_2+\lambda_5)v_\phi\right]+\sqrt{3}c_\alpha\left[8(\lambda_3+\lambda_4)v_\chi+2M_2\right],$$

From above expressions, one can deduce that the loop contributions mainly depend on λ5. To understand why this is so, recall again the expressions:

$$m_5^2 = \frac{M_1}{4v_\chi}v_\phi^2+12M_2v_\chi+\frac{3}{2}\lambda_5v_\phi^2+8\lambda_3v_\chi^2,$$
$$m_3^2 = \frac{M_1}{4v_\chi}(v_\phi^2+8v_\chi^2)+\frac{\lambda_5}{2}(v_\phi^2+8v_\chi^2) = \left(\frac{M_1}{4v_\chi}+\frac{\lambda_5}{2}\right)v^2.$$

λ1 is given by **λ1=(MH²sin²α+mh²cos²α)/8v²ϕ**, λ3 and λ4 are chosen in a narrow window to satisfy the unitarity constraints as shown in figure 7. One then finds that M1 and M2 depend primarily on λ5. λ2 also depends on λ5, through:

**4λ2-λ5= sin2α(mH²-mh²)/2√3vχvϕ+2m²3/v²=5.76±0.56**

The following table summarizes these results, assuming the PDG value **Κγγ = 1.11+0.11-0.09,** which fixes λ5:

| m3 GeV | m5 GeV | sα | sH | λ1/π | λ2/π | λ5/π | λ3/π | λ4/π | M1 GeV | M2 GeV |
|---|---|---|---|---|---|---|---|---|---|---|
| 420 | 660 | -0.15 | 0.5 | 0.02 | 0.01 | -1.7+0.2-0.6 | 0.6 | -0.2 | 1000 | 1000 |

## IV.  An extended GM model

As explained in [1], extending the GM model, EGM, to include a second isodoublet is necessary to interpret the couplings of A(400) to fermions. In short, one has to explain why ATLAS has observed a sizeable signal in A->ττ and why it has observed the most convincing signal for hZ associated with b-jets.

Adding an extra doublet allows tuning independently the couplings of A to τ, b and top.  This extra doublet seems a natural extension of the model but, to my knowledge, has not been tried within the GM phenomenology.

It could also have consequences on g-2 through reinforcing the coupling of A to μ+μ-. In [14] this idea has been tried but with only one triplet and two doublets, meaning that the GM mechanism to protect the ρ parameter is not in action and that one instead assumes a very small vacuum for the triplet component.

In an EGM model doublets and triplets will be a mixture of X and ϕ, with ϕ containing two doublets, hence a CP-odd scalar A2, a charged scalar H2+ and a CP-even scalar H2.

### IV.1 CP-odd sector

It is reasonable to assume that one has:

**A(400)=cosγA3+sinγA2  and A'(x)=-sinγA3+cosγA2**

A(400) will couple to ZH and Zh through the A3 component and to bb and ττ and bb through A2, with the enhanced factors from the Aligned-Two-Higgs-Doublet-Scheme mechanism **A2HDS** [4].

The GM coupling of A3 to tt is ~tanθ$_H$=0.58. The A2 coupling to fermions will go like sinγζfYf. One has:

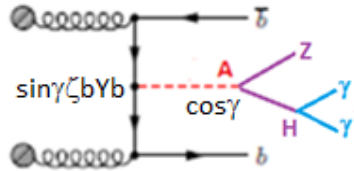

A solution satisfying all the present LHC observation is achieved assuming that A and A' are ~ mass-degenerate and that there is maximal mixing, tanγ~1.

### IV.1 Charged scalar sector

The sector of charged scalars will behave similarly:

**H+= cosγH3$^+$ + sinγH2$^+$ and H'$^+$(x)=-sinγH3$^+$ + cosγH2$^+$**

Γ(H->2γ) will therefore receive an extra contribution from H2$^+$. Note that:

$$C_{H_3^+ H_3^{+*} H} = \frac{1}{\sqrt{3}v^2} \left\{ \sqrt{3}s_\alpha \left[ (4\lambda_2 - \lambda_5)v_\phi^3 + 8(8\lambda_1 + \lambda_5)v_\phi v_\chi^2 + 4M_1 v_\phi v_\chi \right] \right.$$
$$\left. + c_\alpha \left[ 8(\lambda_3 + 3\lambda_4 + \lambda_5)v_\phi^2 v_\chi + 16(6\lambda_2 + \lambda_5)v_\chi^3 + 4M_1 v_\chi^2 - 6M_2 v_\phi^2 \right] \right\}$$

The last term dominates, with M2=1TeV, and gives a large negative coupling of the type -6M2/√3, which adds up to the WW loop and increases Γ(H->2γ).

H2+ could have a similar behaviour which would then further reinforce Γ(H->2γ). Assuming that H3$^+$ and H2$^+$ give the same contribution to H(151), one finds that Γ(H->2γ) can reach ~0.025 MeV, a value sufficient to understand why H(151) is observed into two photons and not into four leptons, a paradox underlined in [2].

These arguments are of course semi-quantitative and one has to treat this problem globally, a future task left to the experts in the field.

### V.     What will happens in e+e- ?

For charged scalars, one will produce H$^+$H$^-$ and H'$^+$H'$^-$, hence an increase in cross section, probably ~twice as large than expected from GM due to the photonic coupling, which will already allow to prove the existence of this extra doublet. Masses could differ allowing further identification. Final states could also differ.

For the neutral scalar sector, H2 decouples from ZZ (two doublet decoupling effect), hence there will be almost no e+e→H2Z contribution but only e+e→ H2A2, with a 800 GeV threshold.

## VI. Lifting mass degeneracies within GM

So far, I have assumed that triplets and 5-plets were mass-degenerate. This hypothesis is unnecessary and can be lifted, which may be of importance for phenomenology. This is shown in [15] where substantial mass differences are predicted. Since then, the authors of this paper have noted the present results deduced from LHC observations and concluded that they are perfectly compatible with their model [16].